\documentstyle[12pt]{article}

\begin{document}
\begin{center}
\Large {\bf Relativistic Quantum Scattering on a Cone}
\end{center}
\centerline{J Spinally, E R Bezerra de Mello and V B Bezerra } 
\begin{center}
Departamento de F\'{\i}sica, Universidade Federal da Para\'{\i}ba\\
Caixa Postal 5008, 58051-970 Jo\~{a}o Pessoa, PB, Brazil.\\
\end{center}

\centerline{\bf Abstract}
\bigskip

In this paper we study the relativistic quantum mechanical scattering
of a bosonic particle by an infinite straight cosmic string, 
considering the non-minimal coupling between the bosonic field and the
scalar curvature. For the idealized cosmic string, an effective 
two-dimensional delta-function interaction takes place besides the usual 
topological scattering. In this case the dimensionless coupling parameter 
must be renomalized in order to make this problem consistent.


\newpage

\section{Introduction}

The non-relativistic classical and quantum scattering of a massive
particle in a locally flat conical two-geometry was developed a few 
years ago by Deser and Jackiw
\cite{Jackiw1}. They showed that there is a topological quantum scattering 
amplitude which presents a singular behavior for scattering angles close 
to $w=\pi(\alpha^{-1}-1)$, where $\alpha$ is a parameter smaller than unity 
which codify the presence of a conical geometry. These results were extended
by allowing the source or the test particle, or both, to carry angular 
momentum\cite{Jackiw2}. In another paper by Gibbons and others\cite{Gibbons},
the non-relativistic scattering problem on a cone was also
analyzed but at this time considering a charged particle and its Coulomb 
self-interaction.

In this paper we analyze the relativistic scattering problem for 
a bosonic spin zero particle in a conical space-time, considering the 
covariant coupling between the scalar field $\psi(x)$ with the 
geometry, the so-called non-minimal coupling. This term, which is an 
invariant one, is expressed by $\xi R\psi$, $\xi$ being a dimensionless 
arbitrary parameter and $R$ the scalar curvature. Because the scalar 
curvature for an idealized cosmic string space-time is proportional to a 
two-dimensional delta-function, this quantum mechanical problem possess 
a non-trivial dynamics: as we shall see, only the s-wave component interacts 
with this delta-potential which indicates that this analysis should be made 
in a more systematic way. 

The analysis of the non-relativistic quantum mechanical problem of a 
particle in the presence of a delta-function potential in two and three 
dimensions has been done by Jackiw \cite{Jackiw}, who pointed out that a 
dimensionless bare coupling constant must be cut-off dependent and that 
it is necessary to introduce a renormalized coupling constant in order 
to make this problem consistent. Also analyzing the fermion-fermion effective 
potential for a Maxwell-Chern-Simons theory \cite{Girotti} 
and a massive Thirring model \cite{Ribeiro} in a (2+1)-dimensional space-time, 
a two-dimensional delta-function potential emerges in both models under 
specific circumstances, and also a systematic approach to provide a 
self-adjoint extension of the respective Hamiltonians operators were performed.
The inclusion in our system of the non-minimal coupling between the scalar 
field and scalar curvature of an idealized cosmic string space-time, as 
far as we know, is the first example where an effective two-dimensional 
delta function interaction naturally occurs in a relativistic quantum 
mechanical problem. Moreover, our analysis has as a purpose to consider the 
scattering problem in a more general case and also to present a 
delta-potential interaction under the geometrical point of view.

This paper is organized as follows. In the Sec. 2, we analyze the 
relativistic scattering problem of a scalar particle by an idealized 
cosmic string considering the non-minimal coupling between this field and 
the scalar curvature. In Sec. 3, we consider the scattering of a charged 
particle, taking into account its repulsive induced self-interaction. As we 
shall see the inclusion of this term produces a complete changing in the 
analysis developed in the previous section. Finally, we leave for Sec. 4 
our conclusions and discussion.  

\section{ Relativistic Quantum Scattering by a \\ Cosmic String}

In this section we study the scattering of a scalar particle in the 
space-time of a cosmic string. We analyze the relativistic scattering
considering a non-minimal coupling in the Klein-Gordon equation, $\xi R\psi$,
between the particle and geometry. The coefficient $\xi$ is an arbitrary 
dimensionless coupling constant and $R$ the scalar curvature.

The line element corresponding to the space-time of a cosmic string is 
given by

\begin{equation}
ds^2=-dt^2+dr^2+\alpha ^2r^2d\theta ^2+dz^2,  \label{6.1}
\end{equation}
where $\alpha =1-4\mu<1$, $\mu$ being the linear mass density of the 
cosmic string. (Here we are considering $G=c=1$).

The Klein-Gordon equation in a general space-time with a non-minimal 
coupling can be written as

\begin{equation}
\left[ \Box ^2-\left( \frac m\hbar \right) ^2-\xi R\right] \psi \left( \vec{r%
},t\right) =0,  \label{6.5}
\end{equation}\\
where $\Box$ is the covariant d'Alembertian operator, and $m$ is the mass of 
the particle. 

In the space-time of a cosmic string described by (\ref{6.1}), the above 
equation turns into

\begin{equation}
\left[ -\frac{\partial ^2}{\partial t^2}+\frac 1r\frac \partial {\partial
r}r\frac \partial {\partial r}+\frac 1{\alpha ^2r^2}\frac{\partial ^2}{%
\partial \theta ^2}+\frac{\partial ^2}{\partial z^2}-\left( \frac m\hbar
\right) ^2-\xi R\right] \psi \left( \vec{r},t\right) =0,  \label{6.6}
\end{equation}
where the scalar curvature is given by $R=R\left( r\right) =
2\left( 1-\alpha \right) \delta \left( r\right)
/\alpha r$.

As the space-time under consideration is static and invariant under 
translations in the $z$ direction, the wave function that solves 
eq.(\ref{6.6}) can be written as

\begin{equation}
\psi _n\left( r,\theta ,z,t\right) =u_n\left( r,\theta \right) \exp \left( -%
\frac{iEt}\hbar \right) \exp \left( ik_zz\right) ,  \label{6.7}
\end{equation}\\
$k_{z}$ being the propagation vector along the $z$ direction and $E$ the 
energy of the particle.

Substituting (\ref{6.7}) into (\ref{6.6}), our problem turns into an 
effective bidimensional one given by the following equation

\begin{equation}
\left[ \frac 1r\frac \partial {\partial r}r\frac \partial {\partial r}+\frac
1{\alpha ^2r^2}\frac{\partial ^2}{\partial \theta ^2}+k^2\right] u_n\left(
r,\theta \right) =V\left( r\right) u_n\left( r,\theta \right),  \label{6.8}
\end{equation}
where $V\left( r\right) =\xi R\left( r\right) $ and
$k^2=\frac{E^2-m^2}{\hbar ^2}-k_z^2.$

The most efficient procedure to deal with a differential equation that 
presents a delta distribution, is to transform it into an 
integral equation, for example, the Lippman-Schwinger equation, which
is the appropriate one in our case. Let us consider that the 
two-dimensional incident wave in the 
conical space can be decomposed by combining Bessel and 
exponential functions as  $J_{\frac{|n|}{\alpha}}(kr)e^{in\theta}$, 
where $n$ is an integer number \cite{Jackiw1}. In this way the 
Lippman-Schwinger equation for the $n$-component wave function becomes 

\begin{eqnarray}
u_n\left( r,\theta \right)&=& J_{\frac{\left| n\right| }\alpha }\left(
kr\right) \exp \left( in\theta \right)
\nonumber \\
                          &&+\alpha \int_0^\infty \int_0^{2\pi}r^{,}dr^{,}
d\theta ^{,}G_n\left( r,\theta ;r^{,},\theta ^{,}\right) V\left(
r^{,}\right) u_n\left( r^{,},\theta ^{,}\right) ,  \label{6.9}
\end{eqnarray}
where $G_n\left( r,\theta ;r^{,},\theta ^{,}\right) $
is the  $n$-th component of the Green function associated with the operator
in the l.h.s. of eq.(\ref{6.8}).

Due to the fact that the solution of the homogeneous equation associated 
with (\ref{6.8}) is given as a product of Bessel and exponential functions, 
we can conclude that the general solution for this equation can be 
obtained considering

\begin{equation}
u_n\left( r,\theta \right) =F_n\left( k;r\right) \exp \left( in\theta \right).
\label{6.11}
\end{equation}
Therefore, substituting (\ref{6.11}) into (\ref{6.9}), we get

\begin{eqnarray}
F_n\left( k;r\right) &=& J_{\frac{\left| n\right| }\alpha }\left( kr\right)
+\alpha \exp \left( -in\theta \right)
\nonumber \\ 
                     & \times &\int_0^\infty \int_0^{2\pi
}r^{,}dr^{,}d\theta ^{,}G_n\left( r,\theta ;r^{,},\theta ^{,}\right) V\left(
r^{,}\right)F_n\left( k;r^{,}\right) 
\exp \left( in\theta ^{,}\right),
\label{6.12}
\end{eqnarray}
where the Green function $G_n\left( r,\theta ;r^{,},\theta ^{,}\right) $
satisfies the following equation

\begin{equation}
\left[ \frac 1r\frac \partial {\partial r}r\frac \partial {\partial r}+\frac
1{\alpha ^2r^2}\frac{\partial ^2}{\partial \theta ^2}+k^2\right] G_n\left(
r,\theta ;r^{,},\theta ^{,}\right) =\frac 1{\alpha r}\delta \left(
r-r^{,}\right) \delta \left( \theta -\theta ^{,}\right) .  \label{6.13}
\end{equation}

Using the representation of the delta function in the angular variable by

\begin{equation}
\delta \left( \theta -\theta ^{,}\right) =\frac 1{2\pi }\sum_n\exp \left[
in\left( \theta -\theta ^{,}\right) \right] ,  \label{6.14}
\end{equation}
we can write the Green function using the {\it ansatz}

\begin{equation}
G\left( r,\theta ;r^{,},\theta ^{,}\right) =\frac 1{2\pi }\sum_n\exp \left[
in\left( \theta -\theta ^{,}\right) \right] g_n\left( r,r^{,}\right),
\label{6.15}
\end{equation}
where $g_{n}\left( r,r^{,}\right)$ is an unknown function which obeys 
the following radial differential equation 
\begin{equation}
\left[ \frac{\partial ^2}{\partial r^2}+\frac 1r\frac \partial {\partial r}-%
\frac{n^2}{\alpha ^2}\frac 1{r^2}+k^2\right] g_n\left( r,r^{,}\right) =\frac
1{\alpha r}\delta \left( r-r^{,}\right) .  \label{6.16}
\end{equation}

The solution of the above equation can be obtained by the standard procedure,
and is given by

\begin{equation}
g\left( r,r^{,}\right) =-\frac{i\pi }{2\alpha }J_{\frac{\left| n\right| }
\alpha }\left( kr_{<}\right) H_{\frac{\left| n\right| }\alpha }^{\left(
1\right) }\left( kr_{>}\right) ,  \label{6.23}
\end{equation}
where $r_{<}$ $\left( r_{>} \right)$ represents the smaller (greater) 
value between $r$ e $r^{,},$ and $H_{\nu}^{\left(1 \right)}\left(z \right)$ 
is the Hankel function.

Finally, substituting (\ref{6.23}) into (\ref{6.15}), we get

\begin{equation}
G\left( r,\theta ;r^{,},\theta ^{,}\right) =\sum_n\left( -\frac i{4\alpha
}\right) \exp \left[ in\left( \theta -\theta ^{,}\right) \right] J_{\frac{%
\left| n\right| }\alpha }\left( kr_{<}\right) H_{\frac{\left| n\right| }%
\alpha }^{\left( 1\right) }\left( kr_{>}\right) ,  \label{6.24}
\end{equation}
and consequently we can write

\begin{equation}
G_n\left( r,\theta ;r^{,},\theta ^{,}\right) =-\frac i{4\alpha }\exp \left[
in\left( \theta -\theta ^{,}\right) \right] J_{\frac{\left| n\right| }\alpha
}\left( kr_{<}\right) H_{\frac{\left| n\right| }\alpha }^{\left( 1\right)
}\left( kr_{>}\right) .  \label{6.25}
\end{equation}
Therefore using the result given by eq.(\ref{6.25}), eq.(\ref{6.12}) turns 
into

\begin{equation}
F_n\left( k;r\right) =J_{\frac{\left| n\right| }\alpha }\left( kr\right)
+i\vartheta \int_0^\infty dr^{,}J_{\frac{\left| n\right| }\alpha }\left(
kr_{<}\right) H_{\frac{\left| n\right| }\alpha }^{\left( 1\right) }\left(
kr_{>}\right) \delta \left( r^{,}\right) F_n\left( k;r^{,}\right) ,
\label{6.26}
\end{equation}
with $\vartheta =\xi \left( 1-\alpha \right) \pi /\alpha .$ (We see 
from (\ref{6.26}) that in the absence of a cosmic string,
$\alpha=1,$ there is no scattering at all.) 

In order to compute the integral in the above expression, let us
divide the interval of integration $[0,\infty )$ as $[0,r>0]\cup
[r>0,\infty )$. Then, we have

\begin{equation}
F_n\left( k;r\right) =J_{\frac{\left| n\right| }\alpha }\left( kr\right)
+i\vartheta H_{\frac{\left| n\right| }\alpha }^{\left( 1\right) }\left(
kr\right) J_{\frac{\left| n\right| }\alpha }\left( 0\right) F_n\left(
k;0\right) .  \label{6.28}
\end{equation}

As the behavior of the Bessel function at the origin 
depends on its order, we have to analyze the  above equation in two 
different cases as follows:\\
i) For $n=0$ ($s$ wave) we have,

\begin{equation}
F_0\left( k;r\right) =J_0\left( kr\right) +i\vartheta H_0^{\left( 1\right)
}\left( kr\right) F_0\left( k;0\right).  \label{6.29}
\end{equation}
ii) For $n\neq 0$

\begin{equation}
F_n\left( k;r\right) =J_{\frac{\left| n\right| }\alpha }\left( kr\right) .
\label{6.30}
\end{equation}

Equations (\ref{6.29}) and (\ref{6.30}) permit us to conclude that
for $n\neq 0$, the wave function does not interact with the 
delta function, so this term does not interfere in the scattering 
process in this case.

Now, let us analyze the case $n=0$ in more detail. Taking $r=\frac 1\Lambda
\rightarrow 0$ into eq.(\ref{6.29}), we have

\begin{equation}
F_0\left(  k;1/\Lambda \right) =
\frac 1{1-\frac \vartheta \pi \left[ 2\ln \left( \frac
\Lambda k\right) +i\pi \right] },  \label{6.31}
\end{equation}
where the regularized expression for 
$H_{0}^{\left( 1\right)}\left( k/\Lambda\right)$ \cite{Abromowitz}
has been used.

From eq.(\ref{6.31}) we conclude that $F_0\left( k;1/\Lambda \right) 
\rightarrow 0$ when $\Lambda\rightarrow \infty $. On the other hand, 
substituting this result into eq.(\ref{6.29}), we find that
$F_0\left( k;r\right) =J_0\left(kr\right),$ which 
in the limit $r\rightarrow 0$, 
gives us $F_0\left( k;0\right) =1 .$ This result contradict the previous
one. As was pointed out in\cite{Jackiw}, this happens because 
of the presence of a 
delta-potential in our effective two-dimensional Schr\"odinger equation,
eq.(\ref{6.8}), which makes the system non-trivial. It is necessary to 
construct a self-adjoint extension of the Hamiltonian operator, $H_{l=0}$, 
in order to have a consistent theory. This procedure is equivalent to 
consider the bare coupling to be cut-off dependent, 
$\vartheta\left( \Lambda \right),$ and to introduce a renormalized coupling 
constant, $\eta\left(M \right).$ In order to do that, let us proceed as 
follows: let us rewrite eq.(\ref{6.31}) in the form

\begin{equation}
\vartheta \left( \Lambda \right) F_0\left( k;1/\Lambda \right) =
\frac 1{\frac 1{\eta
\left( M\right) }-\frac 2\pi \ln \left( \frac Mk\right) -i},  \label{6.32}
\end{equation}
where the renormalized parameter $\eta \left( M \right)$ is defined by 

\begin{equation}
\frac 1{\eta \left( M\right) }=\frac 1{\vartheta \left( \Lambda \right)
}-\frac 2\pi \ln \left( \frac \Lambda M\right).  \label{6.33}
\end{equation}
Here $M$ is a subtraction point. Let us assume also that 
$\vartheta \left( \Lambda \right) \rightarrow 0$ when
$\Lambda \rightarrow \infty ,$ in such way that the right hand side of
eq.(\ref{6.33}) remains finite. We also assume that 
$F_{0}\left( k;1/\Lambda \right) \rightarrow \infty$ in this limit, and 
that the product $\vartheta\left( \Lambda\right)F_{0}\left( k;
1/\Lambda\right)$ is finite. Therefore, from (\ref{6.29}), we find 

\begin{equation}
F_0\left( k;r\right) =J_0\left( kr\right) +i\frac{H_0^{\left( 1\right) }\left(
kr\right) }{\frac 1{\eta \left( M\right) }-\frac 2\pi \ln \left( \frac
Mk\right) -i},  \label{6.34}
\end{equation}
and as a consequence this incompatibility is ruled out.

Notice that in this regularization procedure, the dimensionless coupling 
parameter $\xi$ becomes cut-off dependent. In fact this parameter is related 
with $\vartheta$ by $\vartheta \left( \Lambda \right)=
\xi \left( \Lambda \right) \left(1-\alpha \right)\pi/\alpha.$

The coupling term that appears into eq.(\ref{6.5}), can also be 
interpreted as a correction to the mass of the particle.
In this way, eq.(\ref{6.5}) can be viewed as the equation that
describes the relativistic quantum  behavior of a free particle in the 
conical space-time with effective mass $m\left( r\right) =
\left( m^2+\xi \left(\Lambda \right) R\left( r\right) \right) ^{\frac 12}$,

Now, let us study the scattering problem in this geometry and show up the 
contribution of the non-minimal coupling term. In accordance with 
eq.(\ref{6.7}), the complete solution of eq.(\ref{6.6}) is given by

\begin{equation}
\psi \left( r,\theta ,z,t\right) =\sum_na_nu_n\left( r,\theta \right) \exp
\left( -\frac{iEt}\hbar \right) \exp \left( ik_zz\right) ,  \label{6.35}
\end{equation}
where $a_n$ is a constant.
As the energy of the scattered particle does not depend on the quantum
number $n$, we can write eq.(\ref{6.35}) as

\begin{equation}
\psi \left( r,\theta ,z,t\right) =\exp \left( -\frac{iEt}\hbar \right) \exp
\left( ik_zz\right) \sum_na_nu_n\left( r,\theta \right) .  \label{6.36}
\end{equation}

Therefore, we have

\begin{eqnarray}
u\left( r,\theta \right) &=&\sum_na_nu_n\left( r,\theta \right)  \nonumber \\
&=&\sum_na_n J_{\frac{\left| n\right| }\alpha }\left( kr\right)
\nonumber \\
&+&i\vartheta \left( \Lambda \right)\sum_na_n H_{\frac{\left| n\right| }\alpha
}^{\left( 1\right) }\left( kr\right) J_{\frac{\left| n\right| }\alpha
}\left( 0\right) F_n\left( k;0\right)\exp \left( in\theta \right) .
\label{6.37}
\end{eqnarray}

On the other hand, using the standard procedure we have

\begin{equation}
u\left( r,\theta \right) _{\overrightarrow{r\rightarrow \infty }}\exp \left[
ikr\cos \left( \theta \right) \right] +\sqrt{\frac ir}f\left( \theta \right)
\exp \left( ikr\right) ,  \label{6.38}
\end{equation}

or

\begin{eqnarray}
u\left( r,\theta \right)_{\overrightarrow{r\rightarrow \infty }%
}&&\sum_n\left( i\right) ^n\sqrt{\frac 2{\pi kr}}\cos \left( kr-\frac{n\pi }%
2-\frac \pi 4\right) \exp \left( in\theta \right) 
\nonumber \\
&&+\sqrt{\frac ir}f\left(
\theta \right) \exp \left( ikr\right) ,  \label{6.39}
\end{eqnarray}
where $f\left( \theta \right) $ is the scattering  amplitude.

Let us consider the asymptotic behavior of eq.(\ref{6.37}) which is given by

\begin{eqnarray}
&&u\left( r,\theta \right) _{\overrightarrow{r\rightarrow \infty }}\sum_na_n%
\sqrt{\frac 2{\pi kr}}\{\cos \left( kr-\frac{\left| n\right| \pi }{2\alpha }%
-\frac \pi 4\right) +  \nonumber \\
&&i\vartheta \left( \Lambda \right) J_{\frac{\left| n\right| }\alpha }\left(
0\right) F_n\left( 0\right) \exp \left( kr-\frac{\left| n\right| \pi }{%
2\alpha }-\frac \pi 4\right) \}\exp \left( in\theta \right) .  
\label{6.40}
\end{eqnarray}

From results given by eqs.(\ref{6.39}) and (\ref{6.40}), we get

\begin{equation}
a_n=\exp \left( -\frac{i\left| n\right| \pi }{2\alpha }+i\left| n\right| \pi
\right)  \label{6.41}
\end{equation}

and

\begin{eqnarray}
f(\theta) &=&\frac{1}{\sqrt{-2\pi k}}\sum_n\{[\exp
(-i|n| \omega) -1] 
\nonumber \\
&&+i\vartheta(\Lambda)
J_\frac{|n|}{\alpha}(0) F_n(0)
\exp( -i|n| \omega)\} \exp(in\theta) ,  \label{6.42}
\end{eqnarray}
where $\omega =\left( \alpha ^{-1}-1\right) \pi .$ In order to realize the
sum that appears in the first term of the above equation, we will follow
the same procedure used in \cite {Jackiw1}. The second term gives a  
contribution only in the case $n=0$(s-wave). In this way, eq.(\ref{6.42}) 
turns into

\begin{eqnarray}
f\left( \theta \right) &=&\frac{1}{\sqrt{2\pi k}}\{\frac{\sin \left( \omega
\right) }{\cos \left( \omega \right) -\cos \left( \theta \right)}-i\pi %
[\delta \left( \theta -\omega -2\pi n\right) 
\nonumber \\
&&+\delta \left( \theta
+\omega -2\pi n\right) -2\delta \left( \theta -2\pi n\right)] 
+\vartheta \left( \Lambda \right) F_0\left( 0\right) \}.  \label{6.43}
\end{eqnarray}

Considering the portion of the scattered wave that leads to the delta function
contribution in $f(\theta)$ as belonging to the asymptote of a plane wave,
we can move its contribution to the incoming wave, so that we can redefine 
the scattering amplitude as 

\begin{equation}
\tilde{f}\left( \theta \right) =\frac 1{\sqrt{2\pi k}}\left\{ \frac{\sin
\left( \omega \right) }{\cos \left( \omega \right) -\cos \left( \theta
\right) }+\left[ \frac 1{\eta \left( M\right) }-\frac 2\pi \ln \left( \frac
Mk\right) -i\right] ^{-1}\right\} .  \label{6.44}
\end{equation}

The second term in the above equation is due to the non-minimal coupling $\xi
\left( \Lambda \right) R\left( r\right) .$  It is evident that in the
absence of the coupling term, the scattering amplitude will be given by

\begin{equation}
\tilde{f}\left( \theta \right) =\frac{1}{\sqrt{2\pi k}}\frac{\sin
\omega  }{\cos \omega  -\cos \theta  }, 
\label{6.45}
\end{equation}
which is the same result obtained in\cite{Jackiw1} for the 
non-relativistic case. Note that when $\alpha=1$,
the scattering amplitude vanishes, which is a consistent result.

\section{Relativistic \, Quantum \,Scattering \,of \,a \\ Charged  Particle}

In this section, we want to study the scattering problem of a charged scalar 
particle in the space-time of a cosmic string, taking into account the 
self-interaction term which is given by $V_{ext.}=K/r$, where $r$ is the 
radial distance form the particle to the cosmic string and $K$ a positive 
constant\cite{Linet}. Therefore, in order to solve this problem, we have to 
introduce an additional term into the Klein-Gordon equation corresponding to 
this self-interaction. To do this, there are two different
approaches: the first one considers\cite{Bezerra}, this term as the temporal 
component of the four vector electromagnetic potential $A_{\mu}$. So, the 
self-interaction will produce the following change in the energy 
$E \rightarrow E+V_{ext.}$. The second one, considered in \cite{Bordag}, 
takes this self-energy as a scalar potential. This is equivalent to do a 
correction in the mass term as $m \rightarrow m+V_{ext.}.$ In this paper 
we shall adopt of the second approach in order to simplify our problem. 
Then, following the second approach, the equation that describes the 
behavior of the scalar particle is given by

\begin{eqnarray}
[ -\frac{\partial ^{2}}{\partial t^{2}}+\frac{1}{r}\frac{\partial }{%
\partial r}r\frac{\partial }{\partial r}+
\frac{1}{\alpha^{2}r^{2}}\frac{\partial
^{2}}{\partial \theta ^{2}}+\frac{\partial ^{2}}{\partial z^{2}}-\left( 
\frac{m}{\hbar }\right) ^{2}-\frac{2mK}{r}
\nonumber \\
-\frac{K^{2}}{r^{2}}+\xi R]
\psi _{n}\left( \vec{r},t\right) =0.  \label{30.1}
\end{eqnarray}

Using the same arguments of the previous section, we can write the solution
of the above equation as

\begin{equation}
\psi _{n}\left( r,\theta ,z,t\right) =P_{n}\left( r,\theta \right) \exp
\left( -\frac{Et}{\hbar }\right) \exp \left( ik_{z}z\right) .  \label{30.2}
\end{equation}

Substituting (\ref{30.2}) into (\ref{30.1}), we get the following 
differential equation for the function $P_{n}\left( r,\theta \right) $

\begin{equation}
\left[ \frac{1}{r}\frac{\partial }{\partial r}r\frac{\partial }{\partial r}+%
\frac{1}{\alpha^{2}r^{2}}\frac{\partial ^{2}}{\partial \theta ^{2}}+k^{2}-
\frac{%
2mK}{r}-\frac{K^{2}}{r^{2}}\right] P_{n}\left( r,\theta \right) =V\left(
r\right) P_{n}\left( r,\theta \right)  \label{30.3}
\end{equation}
where $V\left( r\right) =-\xi R\left( r\right) $ and 
$k^{2}=\frac{E^{2}-m^{2}}{%
\hbar ^{2}}-k_{z}^{2}.$

In order to obtain the solution of eq.(\ref{30.3}), we shall adopt the 
previous procedure and write $P_{n}\left( r,\theta \right)$ as

\begin{eqnarray}
P_{n}\left( r,\theta \right)&=&\frac{M_{\sigma ,\gamma }\left( 2ikr\right) }{%
\sqrt{r}}\exp \left( in\theta \right) 
\nonumber \\
&&+\alpha\int_{0}^{\infty }\int_{0}^{2\pi
}r^{,}dr^{,}d\theta ^{,}G_{n,K}\left( r,\theta ;r^{,},\theta ^{,}\right)
V\left( r^{,}\right) P_{n}\left( r^{,},\theta ^{,}\right) .  \label{30.4}
\end{eqnarray}

The solution of the homogeneous equation associated with eq.(\ref{30.3}), 
which is regular at the origin is the Whittaker function 
$M_{\sigma ,\gamma _{n}}\left( 2ikr\right)$, with $\sigma =\frac{imK}{k}$ 
and $\gamma _{n}=\sqrt{\frac{n^{2}}{ \alpha^{2}}+K^{2}}.$ 
The function that appears in eq.(\ref{30.3}),
$G_{n,K}\left( r,\theta ;r^{,},\theta ^{,}\right)$ is the $n$-th 
component of the Green function, 
$G_{K}\left( r,\theta ;r^{,},\theta \right),$ , associated with the operator
in the l.h.s. of eq.(\ref{30.3}), and must satisfies the following  equation

\begin{eqnarray}
\left[ \frac{1}{r}\frac{\partial }{\partial r}r\frac{\partial }{\partial r}+
\frac{1}{\alpha^{2}r^{2}}\frac{\partial ^{2}}{\partial \theta ^{2}}+
k^{2}-\frac{2mK}{r}-\frac{K^{2}}{r^{2}}\right] G_{K}\left( r,\theta ;
r^{,},\theta^{,}\right) 
\nonumber \\
=\frac{1}{\alpha r}\delta \left( r-r^{,}\right) 
\delta \left( \theta
-\theta ^{,}\right) .  \label{30.8}
\end{eqnarray}

Writing the solution of eq.(\ref{30.3}) as

\begin{equation}
P_{n}(r, \theta) = H_{n}(k;r)exp(in \theta), \label{30.6}
\end{equation}
and substituting into (\ref{30.4}), we get

\begin{eqnarray}
H_{n}\left( k;r\right)&=&\frac{M_{\sigma ,\gamma _{n}}\left( i2kr\right) }{%
\sqrt{r}}+\alpha e^{-in\theta }
\int_{0}^{\infty }\int_{0}^{2\pi
}r^{,}dr^{,}d\theta ^{,}G_{n,K}\left( r,\theta ;r^{,},\theta ^{,}\right)
\{V\left( r^{,}\right) 
\nonumber \\
&& \times H_{n}\left( k;r^{,}\right) \exp \left( in\theta
^{,}\right)\} .  \label{30.7}
\end{eqnarray}

Using (\ref{6.14}) to represent the delta function in the angular variable, 
we also can write  the Green function $G_{K}$ as

\begin{equation}
G_{K}\left( r,\theta ;r^{,},\theta ^{,}\right) =\frac{1}{2\pi }\sum_{n}\exp %
\left[ in\left( \theta -\theta ^{,}\right) \right] g_{n,K}\left(
r,r^{,}\right) .  \label{30.10}
\end{equation}
Putting eq.(\ref{30.10}) into eq.(\ref{30.8}), we have

\begin{equation}
\left[ \frac{d^{2}}{dr^{2}}+\frac{1}{r}\frac{d}{dr}-\left( \frac{\gamma
_{n}^{2}}{r^{2}}+\frac{2mK}{r}-k^{2}\right) \right] g_{n,K}\left(
r;r^{,}\right) =\frac{1}{\alpha r}\delta \left( r-r^{,}\right) .  \label{30.11}
\end{equation}

We can see that (\ref{30.11}) reduces to (\ref{6.16}) in the limit $K=0$. 
Taking into account this fact, we can write the function 
$g_{n,K}\left( r;r^{,} \right)$ in such a way that it reduces to 
(\ref{6.23}), when we take $K=0$. Therefore, we can express this
function in the following way: when $r\neq r^{,},$ the solution of the above
equation is given by $g_{n,K}^{\left( I\right) }\left( r,r^{,}\right)=
A_{n,K}\frac{1}{\sqrt{r}}M_{\sigma ,\gamma _{n}}\left( i2kr\right) $ 
($r<r^{,}$) and $g_{n,K}^{\left(II\right) }\left( r,r^{,}\right)=
B_{n,K}\frac{1}{\sqrt{r}}\tilde{W}_{\sigma,\gamma _{n}}\left( i2kr\right) $
($r>r^{,}$), where $M_{\sigma ,\gamma_{n}}\left( i2kr\right) $ is the usual
Whittaker function, and the function $\tilde{W}_{\sigma ,\gamma _{n}}
\left( i2kr\right)$ is a function appropriately constructed in such that
in the limit $K\rightarrow 0,$ we get the function 
$B_{n}H_{\frac{|n|}{\alpha}}^{(1)}(kr)$. The function 
$\tilde{W}_{\sigma ,\gamma _{n}}\left( i2kr\right)$ is given by

\begin{eqnarray}
\tilde{W}_{\sigma ,\gamma _{n}}\left( i2kr\right) &=&- \frac{\Gamma
\left( -2\gamma _{n}\right) }{\Gamma \left( \frac{1}{2}-\gamma _{n}-K\right) 
}e^{-i2\pi \gamma _{n}}M_{\sigma ,\gamma _{n}}\left( i2kr\right) 
\nonumber \\
&&-\frac{\Gamma \left( 2\gamma _{n}\right) }{\Gamma \left( \frac{1}{2}+\gamma
_{n}-K\right) }M_{\sigma ,-\gamma _{n}}\left( i2kr\right) 
\label{30.12}
\end{eqnarray}

We can determine the constants $A_{n,K}$ and $B_{n,K}$, using the 
continuity of the function $g_{n,K}\left( r,r^{,}\right) $ at 
$r=r^{,}$ and the discontinuity of the first derivative at this point. The
continuity of $g_{n,K} \left( r,r^{,}\right) $ at $r=r^{,}$ gives us

\begin{equation}
A_{n,K}M_{\sigma ,\gamma _{n}}\left( i2kr^{,}\right) =B_{n,K}\tilde{W}%
_{\sigma ,\gamma _{n}}\left( i2kr^{,}\right) .  \label{30.13}
\end{equation}

Integrating (\ref{30.11}) between $r^{,}-\varepsilon $ and $%
r^{,}+\varepsilon ,$ we find the discontinuity of $g_{n,K}\left(
r,r^{,}\right) $ at $r=r^{,}$, which is given by

\begin{equation}
\left[ \frac{\partial }{\partial r}g_{n,K}^{\left( II\right) }\left(
r,r^{,}\right) -\frac{\partial }{\partial r}g_{n,K}^{\left( I\right) }\left(
r,r^{,}\right) \right] _{r=r^{,}}=\frac{1}{\alpha r^{,}}.  \label{30.14}
\end{equation}

Therefore, substituting expressions for $g_{n,K}^{\left( I\right)
}\left( r,r^{,}\right) $ and $g_{n,K}^{\left( II\right) }\left( r,r^{,}\right) 
$ into (\ref{30.14}) and using (\ref{30.13}), we get the following equation

\begin{equation}
\left. B_{n,K}\mathcal{W}\right| _{r=r^{,}}=\frac{1}{\alpha r^{,}}
\frac{1}{\sqrt{r^{,}}}
M_{\sigma ,\gamma }\left( i2kr^{,}\right) ,  \label{30.15}
\end{equation}
where $\mathcal{W}$ is the Wronskian which is given by

\begin{eqnarray}
{\mathcal{W}}(r)&=&\frac{1}{\sqrt{r}}M_{\sigma ,\gamma _{n}}
\left( i2kr\right) 
\frac{d}{dr}\left[ \frac{1}{\sqrt{r}}\tilde{W}_{\sigma ,\gamma _{n}}\left(
i2kr\right) \right] 
\nonumber \\
&&-\frac{1}{\sqrt{r}}\tilde{W}_{\sigma ,\gamma _{n}}\left(
i2kr\right) \frac{d}{dr}\left[ \frac{1}{\sqrt{r}}M_{\sigma ,\gamma
_{n}}\left( i2kr\right) \right] .  \label{30.16}
\end{eqnarray}
 
In order to compute $\mathcal{W}$, we shall use the asymptotic expressions
for $M_{\sigma ,\gamma _{n}}\left( i2kr\right) $ and $\tilde{W}_{\sigma ,\gamma
_{n}}\left( i2kr\right)$. These functions behave, near the origin as
\cite{Abromowitz}:

\begin{eqnarray}
&&\left. \tilde{W}_{\sigma ,\gamma _{n}}\left( i2kr\right) _
{\overrightarrow{
r\rightarrow 0}}-\frac{\Gamma \left( 2\gamma _{n}\right) }{\Gamma \left( 
\frac{1}{2}+\gamma _{n}-i\frac{mK}{k}\right) }\left( i2kr\right) ^{\frac{1}
{2}\gamma _{n}}\right.  \label{30.17} 
\end{eqnarray}
and
\begin{eqnarray}
&&\left. M_{\sigma ,\gamma _{n}}\left( i2kr\right) _{\overrightarrow{
r\rightarrow 0}}\left( i2kr\right) ^{\frac{1}{2}+\gamma _{n}}\right .
\label{30.18} 
\end{eqnarray}
So, the Wronskian becomes

\begin{equation}
{\mathcal{W}}=i\frac{\Gamma \left( 2\gamma _{n}\right) }
{\Gamma \left( \frac{1}{2}
+\gamma _{n}-i\frac{mK}{k}\right) }\frac{4k\gamma _{n}}{r}.  \label{30.19}
\end{equation}
Substituting the expression for the Wronskian into eq.(\ref{30.15}),
we get
 
\begin{equation}
B_{n,K}=\frac{i}{4k\gamma _{n}\varepsilon \alpha}\frac{1}
{\sqrt{r^{,}}}M_{\sigma
,\gamma _{n}}\left( i2kr^{,}\right)  \label{30.20}
\end{equation}
and as a consequence
\begin{equation}
A_{n,K}=\frac{i}{4k\gamma _{n}\varepsilon \alpha}\frac{1}{\sqrt{r^{,}}}\tilde{W}
_{\sigma ,\gamma _{n}}\left( i2kr^{,}\right) ,  \label{30.21}
\end{equation}
where $\varepsilon =-\frac{\Gamma \left( 2\gamma _{n}\right) }{\Gamma \left( 
\frac{1}{2}+\gamma _{n}-i\frac{mK}{k}\right) }.$ Therefore,
$g_{n,K}\left( r;r^{,}\right)$ is given by

\begin{equation}
g_{n,K}\left( r;r^{,}\right) =\frac{i}{4k\gamma _{n}\varepsilon \alpha}\frac{1}{
\sqrt{r_{>}r_{<}}}\tilde{W}_{\sigma ,\gamma _{n}}\left( i2kr_{>}\right)
M_{\sigma ,\gamma _{n}}\left( i2kr_{<}\right) .  \label{30.22}
\end{equation}
Finally, the $n$-th component of the Green function, $G{n,K}$, is 
\begin{equation}
G_{n,K}=\frac{i}{8\pi k\gamma _{n}\varepsilon \alpha}e^{in\left( \theta -\theta
^{,}\right) }\frac{1}{\sqrt{r_{>}r_{<}}}\tilde{W}_{\sigma ,\gamma
_{n}}\left( i2kr_{>}\right) M_{\sigma ,\gamma _{n}}\left( i2kr_{<}\right) .
\label{30.23}
\end{equation}

It is important to call attention to the fact that in the limit
$K\rightarrow 0,$ this Green function coincides with the 
previous one given by eq.(\ref{6.25}). Using this Green function and the 
same procedure of the previous section, we have from eq.(\ref{30.7}) that
\begin{eqnarray}
H_{n}\left( k;r\right) &=&\frac{1}{\sqrt{r}}M_{\sigma ,\gamma _{n}}\left(
i2kr\right)-\frac{i\left( 1-\alpha \right) \xi }{2k\varepsilon \gamma _{n} 
\alpha}\frac{1}{%
\sqrt{r}}\tilde{W}_{\sigma ,\gamma _{n}}\left( i2kr\right)
\nonumber \\
& \times& \int_{0}^{\infty
}dr^{,}\frac{1}{\sqrt{r}}M_{\sigma ,\gamma _{n}}\left( i2kr\right)
H_{n}\left( k;r^{,}\right) \delta \left( r^{,}\right) .  \label{30.24}
\end{eqnarray}

Writing the Whittaker function in terms of the hyper-geometric 
confluent function and assuming that the function $H_{n}\left( k;r\right)$ 
is well behaved at the origin, we can show that the second term of the
above equation vanishes. Therefore, the radial part of the wave function
$H_{n}\left( k;r\right)$ reduces to

\begin{equation}
H_{n}\left( k;r\right) =\frac{1}{\sqrt{r}}M_{\sigma ,\gamma _{n}}\left(
i2kr\right) .  \label{30.25}
\end{equation}

So, we can conclude that the delta function potential which appears into the 
Klein-Gordon equation due to non-minimal coupling does not
contribute to the scattered wave function $\psi$. This occurs due to the 
fact that the self-interaction suppress the possibility of the 
particle be at the origin, even in the case when self-interaction coupling 
constant is very small. Assuming that $E^{2}>m^{2},$ let us 
analyze the scattering process. To do this, let us write  
$H_{n}\left( k;r\right)$ in terms of the 
hyper-geometric confluent function $\Phi \left( a,b;i2kr\right) $ 

\begin{equation}
H_{n}\left( k;r\right) =\frac{1}{\sqrt{r}}e^{ikr}\left( i2kr\right) ^{\frac{1}{
2}+\gamma _{n}}\Phi \left( a,b;i2kr\right) ,  \label{30.26}
\end{equation}
where $a=\frac{1}{2}+\gamma _{n}-i\frac{mK}{k}$ and $b=1+2\gamma _{n}$. Using
the asymptotic behavior of $\Phi \left( a,b;z\right), $ 
given by\cite{Abromowitz}
\begin{equation}
\Phi \left( a,b;z\right) _{\overrightarrow{\left| z\right| 
\rightarrow \infty}}%
\frac{\Gamma \left( b\right) \left( -z\right) ^{-a}}{\Gamma \left(
b-a\right) }+\frac{\Gamma \left( b\right) }{\Gamma \left( a\right) }%
e^{z}z^{a-b},  \label{30.27}
\end{equation}
we get the expression for the asymptotic behavior of the function
$H_{n}\left(k;r \right)$

\begin{equation}
H_{n}\left( k;r\right) _{\overrightarrow{r\rightarrow \infty}}
\frac{1}{\sqrt{r}}
\cos \left[ kr-\frac{mK}{k}\ln \left( 2kr\right) +\eta _{n}-\frac{\pi }{2}
\left( \frac{1}{2}+\gamma _{n}\right) \right] ,  \label{30.28}
\end{equation}
where $\eta _{n}=\arg \Gamma \left( \frac{1}{2}+\gamma _{n}+i\frac{mK}{k}
\right) .$
From the above equation, we can obtain the phase shift $\delta _{n}$, 
which is given by

\begin{equation}
\delta _{n}=\eta _{n}+\frac{\pi }{2}\left( n-\gamma _{n}\right) .
\label{30.29}
\end{equation}
Following the usual procedure, we get the following expression for the 
scattering amplitude 
\begin{equation}
f\left( \theta \right) =\frac{1}{\sqrt{-2\pi k}}\sum_{n}\left( e^{i\left[
2\eta _{n}+\pi \left( n-\gamma _{n}\right) \right] }-1\right) e^{in\theta }.
\label{30.30}
\end{equation}

As we can see, the scattering amplitude depends on the topology of the
space-time and also on the self-interaction coupling constant. 
When we take $K=0$ in this expression, we re-obtain the scattering amplitude 
computed in \cite{Jackiw1}. An unexpected fact is that we do not get the 
result of the previous section in this limit, given by eq.(\ref{6.44}). 
This is associated with the fact that the self-interaction term suppress 
the effect of the delta function, and if we take the limit $K=0$ after the
integration in eq.(\ref{30.24}), the final result vanishes. On the other
hand, taking this limit before integration, our result would agree with the
one given by eq.(\ref{6.44}).

\section{Concluding Remarks}

In this paper we have analyzed the relativistic quantum scattering problem 
of a scalar particle by an idealized infinity straight cosmic string, take 
into account a non-minimal coupling between the bosonic particle and the 
geometry. In this context we have an effective two-dimensional system where 
naturally a delta-potential emerges. The way to deal with this problem is 
to consider the dimensionless coupling as a cut-off dependent parameter, 
$\xi\left(\Lambda \right),$ in such way that 
$\xi\left(\Lambda \right)_{\overrightarrow{\Lambda \rightarrow \infty}}0$, 
and to introduce a renormalized coupling, $\xi\left(M \right)$, which is 
finite and depend on some renormalized point $M$. As a consequence the 
scattering amplitude acquires a logarithm dependence. 

Considering an electrically charged particle in this cosmic string space-time, 
we have show that the situation is completely modified by the presence of the 
repulsive self-interaction. In this case the particle never reach the cosmic 
string, so the delta-interaction potential does not contribute to the 
scattering problem.

Finally, we want to make a brief comment about the inclusion of the 
self-interaction in the Klein-Gordon equation in the case of the scattering
of a charged particleSec. If we assume that this self-interaction is the 
zeroth component of the four-potential $A_{\mu}$, our effective 
two-dimensional Hamiltonian operator will present ill defined solution at 
origin, that is, at the localization of the cosmic string, and therefore,
it is necessary 
to discard this solution in order to have a self-adjoint Hamiltonian 
operator. The others solutions are well defined, vanishing at the origin, 
and so they will not be affected by the delta-potential as well.

\section*{Acknowledgments}

We would like to thank Conselho Nacional de Desenvolvimento
Cient\'\i fico e Tecnol\'ogico (CNPq) and CAPES for partial financial 
support. We are also very grateful to N. R. Khusnutdinov for helpful 
discussion during the course of this work.


\begin{thebibliography}{99}

\bibitem{Jackiw1}  Deser S  and Jackiw R 1988 {\it Commun. Math. Phys.} 
{\bf 118} 495

\bibitem{Jackiw2} P de Sousa Gerbert and R Jackiw  1989 {\it Commun. Math. 
Phys.} {\bf 124} 229

\bibitem{Gibbons} Gibbons G W, Ruiz F R and Vachaspati T 1990 {\it Commun.
Math. Phys.} {\bf 127} 295

\bibitem{Jackiw} Jackiw R 1991 {\it M. A. B. B\'eg Memorial Volume} 
ed A Ali and P Hoodbhoy (Singapore: Word Scientific)

\bibitem{Girotti}  Girotti H O, Gomes M and da Silva A J 1990 
{\it Phys. Lett.} B {\bf 274 } 357

\bibitem{Ribeiro}  Ribeiro R F and Bezerra de Mello E R 1994 {\it Int. J. 
Mod. Phys.} A {\bf 9} 3143

 \bibitem{Abromowitz}  Abromowitz M and Stegun 1972 {\it Handbook 
of Mathematical Function} $9^{th}$ edition (N.Y.: Dover Publications Inc.)

\bibitem{Linet}  Linet B 1986 {\it Phys. Rev.} D {\bf 33} 1833


\bibitem{Bezerra}  Bezerra V B and Araujo I G 1994 {\it Class. Quantum Grav.}
{\bf 11} 1599; Rodrigues Sobreira A A and Bezerra de
Mello E R 1999 {\it Grav. and Cosm.} {\bf 5} 177

\bibitem{Bordag}  Bordag M and Khusnutdinov N R 1996 {\it Class. 
Quantum Grav.} {\bf 13} 41

\end{thebibliography}
\end{document}